\documentclass{emulateapj}
\usepackage{graphicx,multirow}

\renewcommand{\text}[1]{\rm{#1}}

\slugcomment{Draft Version December 18, 2014.}
\shorttitle{Pulse amplitude depends on kHz QPO frequency}
\shortauthors{Bult et al.}

\begin{document}

\title{Pulse amplitude depends on kHz QPO frequency in the accreting millisecond pulsar SAX~J1808.4--3658}

\author{Peter Bult and Michiel van~der~Klis}
\affil{Anton Pannekoek Institute,University of Amsterdam, Postbus 94249, 1090 GE Amsterdam, The Netherlands}
\email{p.m.bult@uva.nl}

\begin{abstract}
We study the relation between the 300--700 Hz upper kHz quasi-periodic oscillation (QPO) and the 401~Hz coherent pulsations across all outbursts of the accreting millisecond X-ray pulsar SAX~J1808.4--3658 observed with the {\it Rossi X-ray Timing Explorer}. We find that the pulse amplitude systematically changes by a factor of $\sim2$ when the upper kHz QPO frequency passes through 401~Hz: it halves when the QPO moves to above the spin frequency and doubles again on the way back.  This establishes for the first time the existence of a direct effect of kHz QPOs on the millisecond pulsations and provides a new clue to the origin of the upper kHz QPO. We discuss several scenarios and conclude that while more complex explanations can not formally be excluded, our result strongly suggests that the QPO is produced by azimuthal motion at the inner edge of the accretion disk, most likely orbital motion.  Depending on whether this azimuthal motion is faster or slower than the spin, the plasma then interacts differently with the neutron-star magnetic field. The most straightforward interpretation involves magnetospheric centrifugal inhibition of the accretion flow that sets in when the upper kHz QPO becomes slower than the spin. 
\end{abstract}


%
% Main body
%
\section{Introduction}
    The low-mass X-ray binary SAX~J1808.4--3658 (SAX~J1808) is a 401 Hz
    accreting millisecond X-ray pulsar \citep[AMXP;][]{Wijnands1998} with a
    2 hr orbital period \citep{Chakrabarty1998}. The system is an X-ray
    transient and has shown five X-ray outbursts between 1998 and 2011
    \citep{Hartman2008,Patruno2012} that were observed with the {\it Rossi
    X-ray Timing Explorer} ({\it RXTE}). The pulsations of SAX~J1808 are
    near-sinusoidal with amplitudes of 2--7\% \citep{Hartman2008,Hartman2009,
    Patruno2012}. They are produced by thermal hotspots, emission regions near
    the stellar magnetic poles, undergoing periodic aspect variations due to
    the stellar spin \citep{Davidson1973,Ghosh1978}. The hotspots, which may
    cover a large fraction of the surface \citep{Lamb2009}, are heated by the
    impact of plasma channeled by the magnetic field towards the poles. The
    pulsations therefore offer a direct probe of the inner accretion flow and
    specifically of the magnetic threading of the accretion disk.

    SAX~J1808 was the first AMXP discovered \citep{Wijnands1998}, and also the
    first pulsar to show the twin kHz quasi-periodic oscillations \citep[QPOs;][]{Wijnands2003} now known to be
    ubiquitous among accreting low-magnetic field neutron stars
    \citep{Klis2006}.  In SAX~J1808 usually only the higher-frequency of the
    twin peaks is observed \citep{Wijnands2003, Straaten2005}, at frequencies
    up to 700~Hz, and it is this ÔupperÕ kHz QPO we report on here. The lower
    kHz QPO is seen only on rare occasions \citep{Wijnands2003}.

    Some kHz QPO models explain the upper kHz QPO with orbital motion of
    short-lived inhomogeneities at some preferred radius in the accretion flow
    \citep{Klis1996, Strohmayer1996, Miller1998, Stella1999, Kluzniak2004, 
    Alpar2008}, others invoke alternative mechanisms \citep{Lai1998, Kato2004, 
    Zhang2004, Bachetti2010}, but for lack of observational clues the correct 
    interpretation has remained elusive.

    Although both the plasma channeling responsible for the X-ray pulsations
    and the mechanism producing the kHz QPOs are believed to probe the inner
    accretion flow of AMXPs, no direct relation between pulsations and kHz QPOs
    has been observed so far. In this Letter we show that such a direct
    relation does in fact exists in SAX~J1808. This provides a new clue to the
    origin of the upper kHz QPO and the nature of the inner accretion flow
    towards this AMXP.

\section{Data reduction}
    We consider all pointed observations of SAX~J1808 with {\it RXTE} using the
    Proportional Counter Array \citep{Jahoda2006}. Each observation
    consists of one or multiple continuous 3~ks exposures as set by the data
    gaps associated with the 95~minute satellite orbit.  We use the 16~s
    time-resolution Standard-2 data to create 2--16~keV light curves, which we
    normalize to the Crab \citep[see, e.g,][for details]{Straaten2003}.

    For the stochastic timing analysis we use all GoodXenon and Event data in
    the 2--20~keV energy range. We compute Leahy normalized \citep{Leahy1983}
    power spectra, using 256~s data segments binned to a 1/8192~s
    ($\sim122~\mu$s) time resolution, giving a frequency resolution of
    1/256~Hz and a Nyquist frequency of 4096~Hz. No background subtraction or
    dead-time correction was done before a power spectrum was calculated. The
    resulting power spectra were averaged over intervals of 1--15~ks
    (depending on the number of active PCUs, which was higher for observations
    early in the {\it RXTE} mission) as required to significantly detect the upper
    kHz QPO.

    We subtracted a modeled Poisson noise power spectrum \citep{Zhang1995} from
    the averaged power spectra using the method of \citet{KleinWolt2004}.
    Accounting for the background emission, we then renormalized the power
    spectra to source fractional rms squared per~Hz \citep{Klis1995}.

    We fitted the power spectra with a set of Lorentzian profiles
    \citep{Belloni2002}, with each profile $L_i(\nu | \nu_{0,i}, W_i, r_i)$, a
    function of Fourier frequency $\nu$, described by three parameters: the
    centroid frequency $\nu_{0,i}$, the full-width-at-half-maximum $W_i$, and
    the fractional rms
    \begin{equation}
        r_i = \sqrt{\int_0^\infty L_i(\nu) d\nu}.
    \end{equation}
    The subscript $i$ refers to the specific power spectral component; in
    particular, $L_u$ represents the upper kHz QPO. 

    When its centroid frequency is high, $L_u$ has good coherence and appears
    as a narrow peak in the power spectrum; as the frequency decreases, so does
    the coherence and $L_u$ becomes a broad noise component. We only consider
    the observations for which $L_u$ has good coherence (quality factor $Q
    \equiv \nu_0 / W > 2$).

    For the analysis of the coherent pulsations we use the same data selection
    as for the stochastic analysis. We correct the photon arrival times to the
    Solar System barycenter with the \textsc{ftool faxbary} using the optical
    source position of \citet{Hartman2008}, and subsequently correct the data
    for the binary orbital ephemeris \citep{Hartman2008, Hartman2009,
    Patruno2012}. The power spectra of Figure~\ref{fig.pds} were made from data
    corrected in this way, and rebinned such that the pulse fundamental
    occupies a single frequency bin. In our coherent analysis of the
    pulsations, we fold the data on the pulse period \citep{Hartman2008} and
    for each pulse waveform measure the sinusoidal amplitude\footnote{Note that
    we report {\it sinusoidal} amplitudes, which are a factor of $\sqrt{2}$
    larger than {\it rms} amplitudes} of the fundamental and second harmonic,
    which we express as a fraction of the mean flux.

\section{Results} 	
\begin{figure}[ht]
		\centering
		\includegraphics[width=\linewidth]{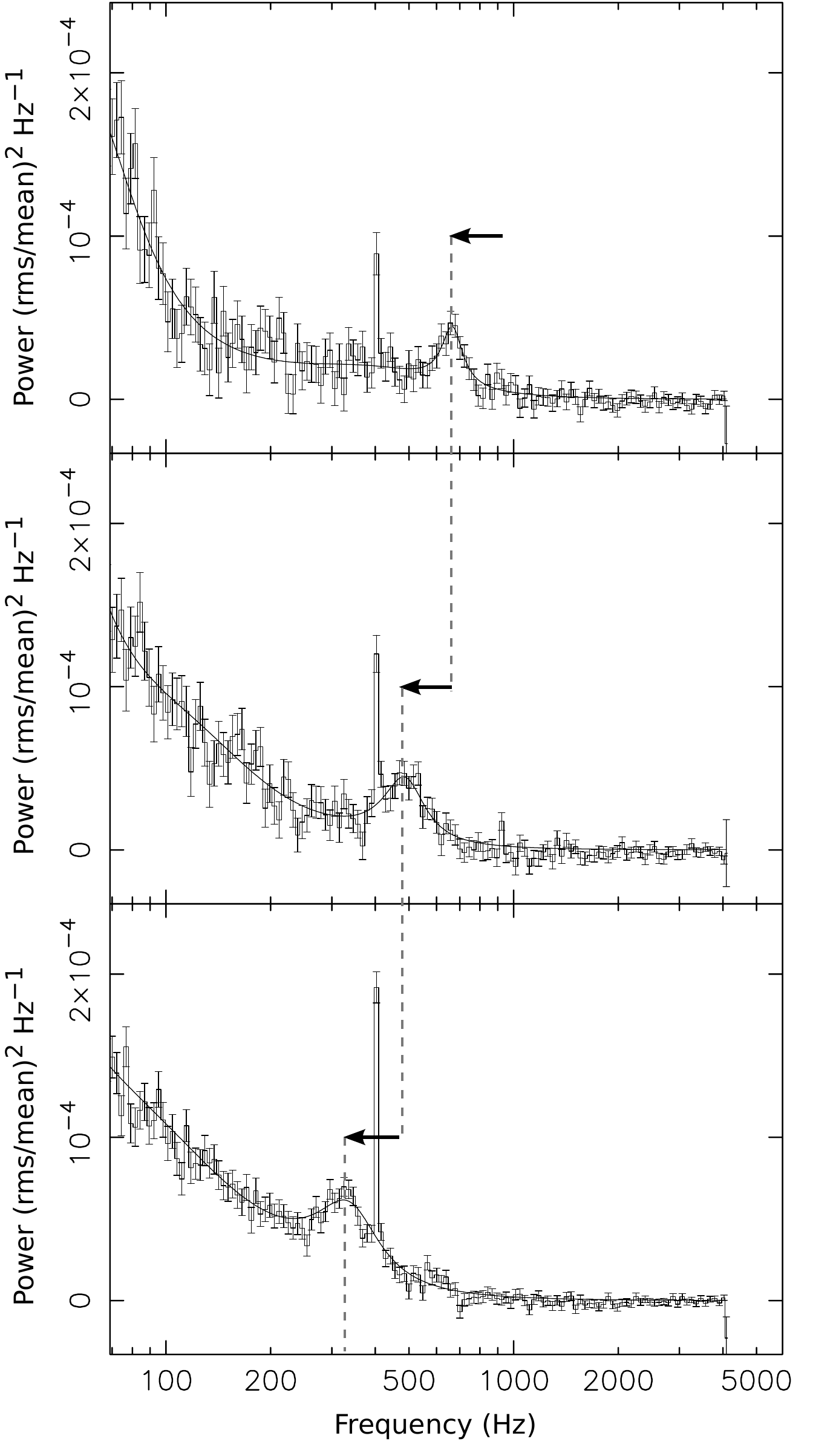}	
		\caption{
            Power spectral evolution during the 2002 outburst, in chronological
            order ({\it top-down}). The frequency evolution of the kilohertz QPO
            (broad peak) in relation to the spike caused by the strictly periodic
            401~Hz pulsations is obvious. When the kHz QPO frequency (dashed lines)
            moves through the spin frequency ({\it second to third frame}) the
            pulse amplitude (height of the spike) increases drastically. Arrows
            indicate the changes in QPO frequency and the initial spike height.
		}
		\label{fig.pds}
	\end{figure}

    In the course of a typical two-week outburst the X-ray flux first rises to
    $\sim70$~milliCrab and then falls back down. In correlation to this, the upper
    kHz QPO drifts from $\sim300$~Hz up to $\sim700$~Hz and back down again, and so
    transits the 401~Hz spin frequency twice per outburst. We find that at these
    transits the amplitude of the pulsations abruptly changes by a factor $\sim2$:
    it halves when in the rise the QPO moves to above the spin frequency and then,
    as illustrated in Figure~\ref{fig.pds}, doubles again on the way back.

		\begin{figure*}[ht]
		\centering
		\includegraphics[width=1.0\linewidth]{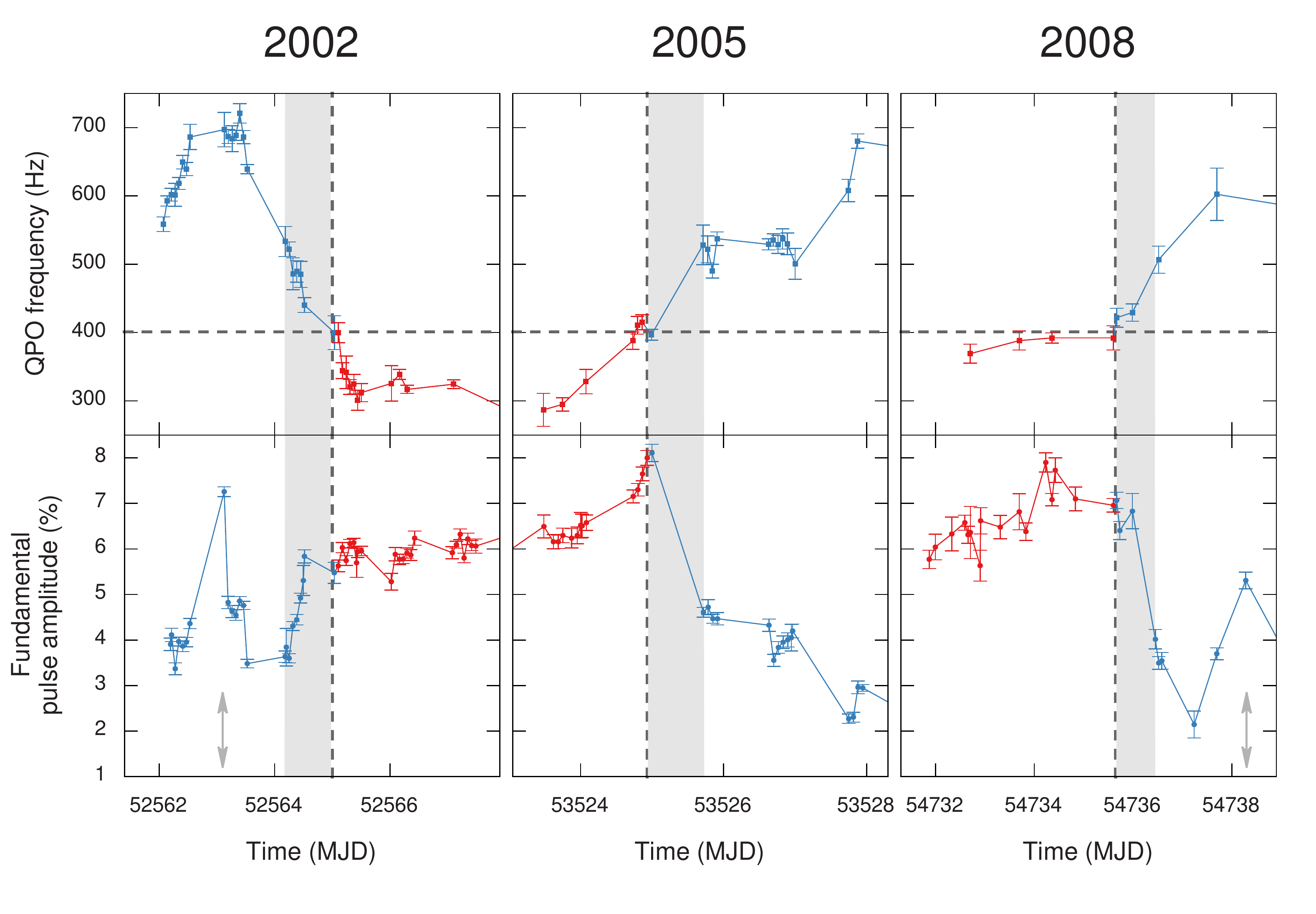}	
		\caption{
            Evolution of the upper kHz QPO frequency ({\it top frame}) and the
            sinusoidal pulse amplitude ({\it bottom frame}) as the QPO
            frequency passes through the 401~Hz spin frequency during outbursts
            in the years indicated. Observations with the QPO frequency above
            and below the spin frequency are marked in blue and red,
            respectively. The dashed lines mark the spin-frequency passages and
            the gray bands indicate the full extent of the transits between the
            two regimes. The double-headed arrows mark pulse amplitude
            excursions (see text).
		}
		\label{fig.transitions}
	\end{figure*}

\begin{deluxetable*}{l c c c c c}
\tabletypesize{\small}
\tablewidth{\linewidth}
\tablecolumns{6}
\tablecaption{ Spin-frequency Transit Measurements. \label{tab.measurements} }
\tablehead{
	\colhead{        } & \colhead{Date } & \colhead{Exposure} & \colhead{QPO frequency} & \colhead{$A_1$} & \colhead{$A_2$} \\
	\colhead{Outburst} & \colhead{(MJD)} & \colhead{(s)     } & \colhead{(Hz)         } & \colhead{(\%) } & \colhead{(\%) }
}
\startdata
\multirow{2}{*}{2002} & 52564.18 & 2800 & $538 \pm 16$ & $3.49 \pm 0.09$ & $0.76 \pm 0.09$ \\
                      & 52564.97 & 3900 & $392 \pm 10$ & $5.63 \pm 0.05$ & $0.63 \pm 0.05$ \\
\hline
\multirow{2}{*}{2005} & 53524.96 & 17500 & $397 \pm 7$ & $7.10 \pm 0.07$ & $1.19 \pm 0.07$ \\
                      & 53525.72 & 13900 & $518 \pm 7$ & $4.35 \pm 0.07$ & $1.35 \pm 0.07$ \\
\hline
\multirow{2}{*}{2008} & 54735.69 & 10700 & $416 \pm 10$ & $6.49 \pm 0.09$ & $0.98 \pm 0.09$ \\
                      & 54736.44 & 12700 & $512 \pm 18$ & $3.33 \pm 0.10$ & $1.34 \pm 0.10$
\enddata
\tablecomments{\ Listing ({\it left-to-right}): the time of observation, total
exposure, upper kHz QPO frequency and the fundamental ($A_1$) and second
harmonic ($A_2$) sinusoidal pulse amplitudes just before and after each
transition for the indicated outbursts. The significances of these changes were
calculated from the errors on the differences using standard error
propagation.}
\end{deluxetable*}

    We observed three such transitions at high signal-to-noise: two where the
    pulse amplitude approximately halves during the outburst rises in 2005 and
    2008, and one where it approximately doubles during the decay in 2002.
    Figure~\ref{fig.transitions} shows the QPO frequency ({\it top frames}) and
    the pulse amplitude ({\it bottom frames}) across these three transitions.
    In each case the pulse amplitude can be seen to systematically drop when
    the QPO moves to frequencies $>401$~Hz ({\it blue points}), and to be
    systematically higher when the QPO is at or below the spin frequency ({\it
    red points}). In 2002, the observations sample the source during the actual
    transition ({\it gray bands}). The pulse amplitude increases steadily from
    3.5 to 5.6\% on a timescale of hours when simultaneously the QPO frequency
    drifts down from $\sim520$ to $\sim400$~Hz. Both amplitude and frequency
    change very significantly ($17\sigma$ and $8\sigma$, respectively, see
    Table~\ref{tab.measurements}). In 2005 and 2008 we observe the opposite
    transition, with behavior that is entirely consistent with that in 2002,
    but time-reversed. The pulse amplitude drops from 7.1 to 4.3\% ($28\sigma$)
    and 6.5 to 3.3\% ($23\sigma$) in 2005 and 2008, respectively, when the QPO
    frequency increases through the 400--520~Hz range ($12\sigma$ and $5\sigma$
    change). In all cases, for QPO frequencies $<401$~Hz the pulse amplitude
    remains at the high level attained near 401~Hz.

\begin{figure}[ht]
		\centering
		\includegraphics[width=1.0\linewidth]{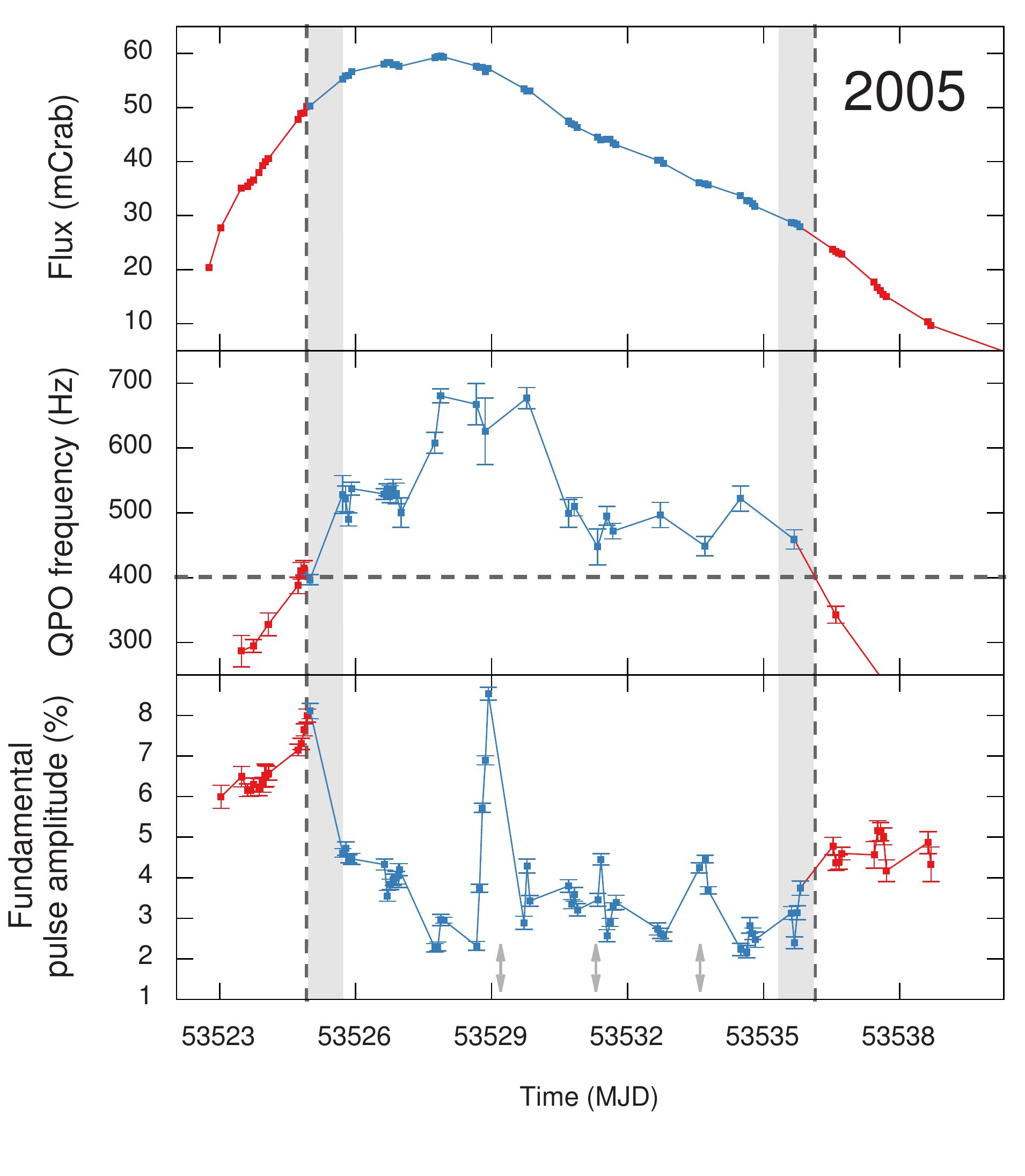}	
		\caption{
            Overview of the 2005 outburst, showing the X-ray flux ({\it top
            frame}), the upper kHz QPO frequency ({\it middle frame}) and the
            fundamental pulse amplitude ({\it bottom frame}) as a function of
            time. Observations with the QPO frequency above and below the spin
            frequency are marked in blue and red, respectively. The dashed
            lines mark the spin-frequency passages and the gray bands show the
            extents of the transits between the two regimes. The double-headed
            arrows mark pulse amplitude excursions (see text). The other
            outbursts show similar trends.
		}
		\label{fig.outburst05}
	\end{figure}

    The pulsations of SAX~J1808 have been studied extensively
    \citep{Hartman2008, Leahy2008, Hartman2009, Ibragimov2009, Kajava2011,
    Morsink2011, Patruno2012} and are known to show brief, hour-timescale
    excursions in amplitude. We define an excursion as a sudden pulse amplitude
    increase of 50\% or more directly followed by a similar amplitude decrease,
    such that the entire episode verifiably takes place within two days. In
    total there are 11 such excursions in the data, and we have marked these in
    Figures~\ref{fig.transitions} and~\ref{fig.outburst05} with double-headed
    gray arrows.

    We find that the occurrence of these brief excursions depends on QPO
    frequency dichotomously. The excursions only occur when the QPOs are
    $>401$~Hz (see, e.g. Figure~\ref{fig.outburst05}). While they are not
    limited to any particular QPO frequency, the most pronounced excursions
    cluster around the highest QPO frequencies of 600--700~Hz.

\begin{figure}[ht]
		\centering
		\includegraphics[width=1.0\linewidth]{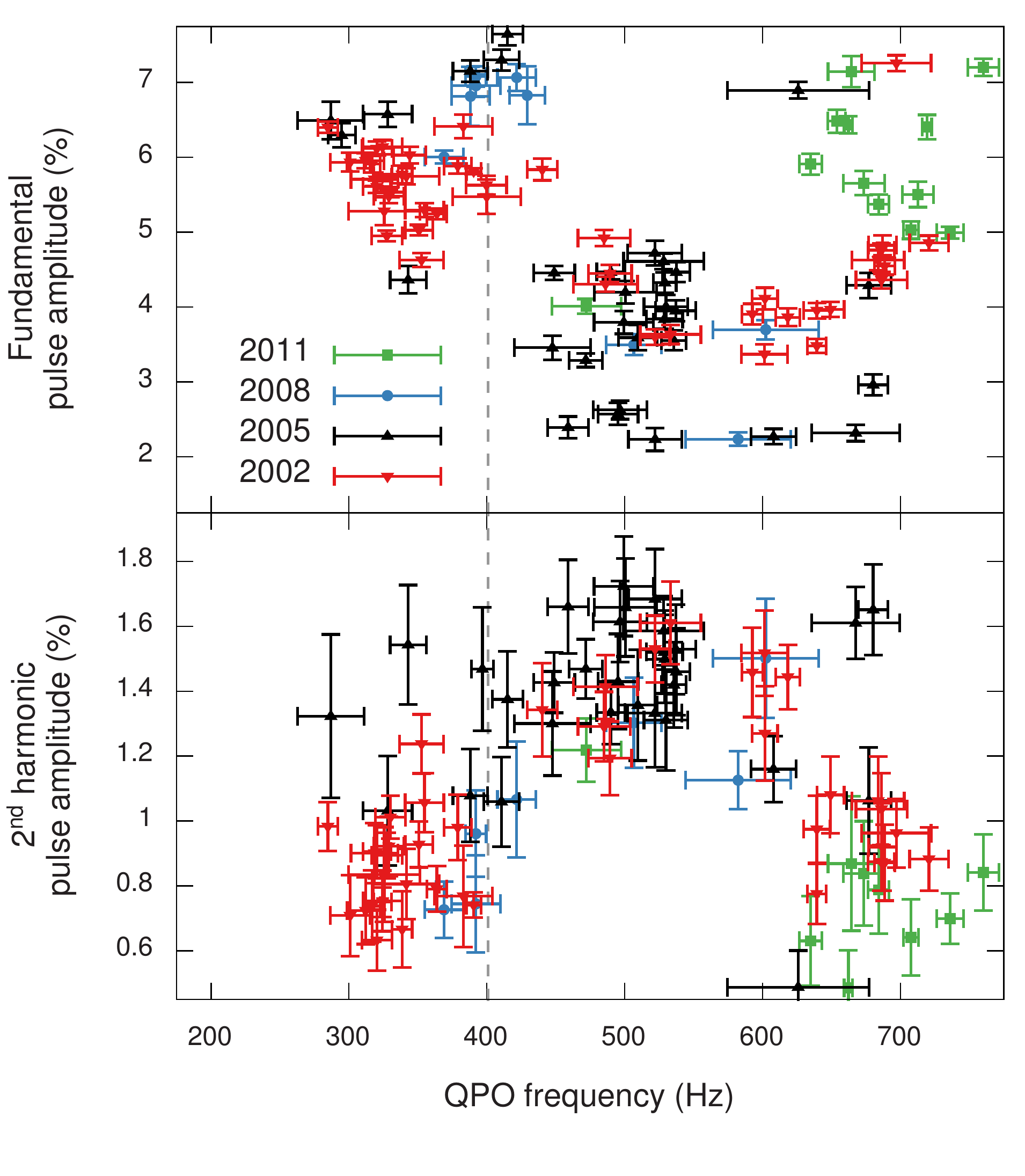}	
		\caption{
            Fundamental ({\it top frame}) and second harmonic ({\it bottom
            frame}) pulse amplitudes as a function of the upper kHz QPO
            frequency of the outbursts as indicated. The dashed line marks the
            spin frequency.
		}
		\label{fig.step}
	\end{figure}

    Figure~\ref{fig.step} shows the amplitudes of the pulse fundamental ({\it
    top frame}) and second harmonic ({\it bottom frame}) versus upper kHz QPO
    centroid frequency as we measured them in all data across all outbursts.
    The large step in pulse amplitude taking place over a QPO frequency range
    of 400--500~Hz is obvious. When in this transition the fundamental
    amplitude drops from $\sim6$~to~$\sim3\%$, the weaker second harmonic
    instead strengthens from $\sim0.9$~to~$\sim1.6\%$. The clustering of the
    brief excursions at high QPO frequencies is also clear among the points
    $>650$~Hz.

    The remaining data, at lower signal-to-noise, show behavior consistent with
    that in the three transitions reported above. When in 2005 at MJD~53536.1
    the QPO moves from above to below 401~Hz, the pulse amplitude increases. In
    1998 there is no transition; the QPO peak, while broad, is always entirely
    below 401~Hz. Consistent with this, the pulsations are strong and the
    second harmonic is weak. In 2011 kHz QPOs are detected only at frequencies
    $>401$~Hz, so no transition is observed. In this outburst the pulse
    amplitude is $\sim4\%$ with, as expected, amplitude excursions of 2--3\% on
    top of this.

\section{Discussion}
    Our analysis establishes, for the first time, a direct effect of kHz QPOs
    on the millisecond pulsations: a large change in pulsation properties
    occurs when the upper kHz QPO frequency transits through the spin
    frequency, but reversely a change in pulse amplitude does not predict the
    QPO frequency. These findings point toward an origin of the upper kHz
    QPO in a process involving azimuthal motion in the accretion flow that
    makes the accreting plasma interact differently with the neutron-star
    magnetic field depending on whether the azimuthal motion is faster or
    slower than the spin. Although spin resonances may play a role (see below),
    mechanisms relying on a resonance of the QPO frequency itself with the spin
    cannot explain our observations: the amplitude of such a resonance is
    largest when the QPO frequency is near 401~Hz and diminishes away from the
    resonant frequency. Instead, what is required is a step-wise change in the
    pulsations between QPO frequencies above and below the spin frequency.

    Our observation that the pulse amplitude is sensitive to whether the QPO
    frequency is faster or slower than the spin immediately suggests an
    interpretation where the QPO is due to orbital motion and the key physics
    involved is the centrifugal force experienced by the accreting plasma as it
    enters the magnetosphere \citep{Illarionov1975, Ghosh1978}. In such a
    scenario, plasma outside the magnetosphere orbits in a Keplerian disk.
    Inside the magnetosphere the motions are dominated by the magnetic field
    and so the plasma is forced to corotate with the stellar spin. As long as the
    radius of the magnetosphere, $r_{\rm m}$, is larger than the
    corotation\footnote{The corotation radius $r_{\rm c}$ is the radius where the
    Keplerian frequency equals the spin frequency: $r_{\rm c}~\equiv~\left[\sqrt{GM}
    / (2\pi\nu_s) \right]^{2/3} = 31$~km for an $M = 1.4 M_\odot, \nu_s =
    401$~Hz star, where $M$ is the neutron-star mass and $\nu_s$ its spin
    frequency.} radius $r_{\rm c}$, the plasma entering the magnetosphere
    speeds up and hence experiences a centrifugal force exceeding gravity,
    inhibiting lateral accretion. For a smaller magnetosphere no such centrifugal 
    inhibition of the lateral accretion flow will occur and therefore a smaller 
    fraction of the accreting plasma is available to flow out of the disk
    plane and toward the poles, causing the pulse amplitude to drop.
    Thus, if the QPO is due to orbital motion at the edge of the magnetosphere
    and hence the QPO frequency can be identified with the Keplerian frequency
    at $r_{\rm m}$, a very different accretion pattern is predicted for QPO
    frequencies above and below the spin frequency. 

    Another possible mechanism that can explain our result can apply to any
    model in which the QPO is due to an azimuthal motion at the inner edge of
    the disk. This motion could be Keplerian, but might also be, e.g.,
    precessional, or that of an azimuthally propagating disk wave \citep[][and
    references therein]{Klis2006}.  As the inner edge of the disk moves inward
    it will pass through a resonance radius, $r_{\rm r}$, where the QPO
    frequency equals the spin frequency and a resonance may occur.  For even
    smaller inner disk radii the QPO frequency increases further, but the
    resonance may continue to exist in the disk at $r_{\rm r}$. It is at least
    logically possible that such a resonance present in the disk flow at radius
    $r_{\rm r}$ leads to a different accretion pattern at radii inside $r_{\rm
    r}$. How precisely the resonance  affects the pulse amplitude remains
    unspecified in this scenario, which in this generic sense could formally
    even work for an azimuthally symmetric QPO mechanism, and/or by interaction
    with the pulsar beam instead of the magnetic field.

    In the above two scenarios, the accretion pattern changes when the inner
    edge of the disk passes through $r_{\rm c}$ or $r_{\rm r}$, respectively. A
    third possibility is that the relevant radius is $r_{\rm m}$. It has been
    suggested that it is possible for the Keplerian disk to penetrate the
    magnetosphere and continue its orbital motion inside $r_{\rm m}$, with the
    inner edge of the disk at $r_{\rm in}$ set by radiative stresses
    \citep{Miller1998}, so that $r_{\rm in}$ can be either larger or smaller
    than $r_{\rm m}$.  If the QPO originates from the inner edge of the disk,
    and its frequency is near the spin frequency as $r_{\rm in}$ passes through
    $r_{\rm m}$, differences in magnetic coupling associated with $r_{\rm in}$
    being larger or smaller than $r_{\rm m}$ might conceivably explain the
    change in the pulsations. For a QPO caused by Keplerian orbital motion
    these conditions might be fulfilled, as during outburst SAX~J1808 may be
    near spin-equilibrium \citep[$r_{\rm m} \simeq r_{\rm c}$,][]{Hartman2008}. 

    So, although some more complex explanations cannot formally be excluded,
    the interpretations discussed here predominantly point to a QPO that is produced
    by azimuthal motion at the inner edge of the accretion disk, most likely
    orbital motion, and a change in the interaction of the accreting plasma
    with the neutron star magnetic field near a QPO frequency equal to the spin
    frequency.  The most straightforward scenario involves centrifugal
    inhibition of the accretion flow that sets in when the upper kHz QPO
    becomes slower than the spin.  This scenario implies that the QPO is
    produced by orbital motion at the edge of the magnetosphere as it passes
    through 401~Hz.

\acknowledgments
	We would like to thank the anonymous referees whose comments improved
	this work. This research has made use of data obtained through the High 
	Energy Astrophysics Science Archive Research Center Online Service, provided 
	by the NASA/Goddard Space Flight Center.

\bibliographystyle{apj}

\end{document}